\RequirePackage{fix-cm}
\documentclass[smallextended]{svjour3}       
\smartqed  
\usepackage{graphicx}
\usepackage{mathptmx}       
\usepackage{helvet}         
\usepackage{courier}        
\usepackage{type1cm}        
\usepackage{url}
\usepackage{makeidx}         
\usepackage{graphicx}        
\usepackage{multicol}        
\usepackage[bottom]{footmisc}

\usepackage{subcaption}

\usepackage[numbers]{natbib}

\begin{document}

\title{Leveraging Evolution Dynamics to Generate Benchmark Complex Networks with Community Structures 
}

\titlerunning{Evolution Dynamics to Generate Networks with Communities}        

\author{Mohammad Qasim Pasta         \and
        Faraz Zaidi 
}


\institute{M. Q. Pasta \at
			Karachi Institute of Economics and Technology \\ 
			and Habib University \\
			\email{qasim.pasta@sse.habib.edu.pk}
           \and
           F. Zaidi \at
           	Karachi Institute of Economics and Technology \\
			\email{faraz.a.zaidi@ieee.org}
}

\date{Received: date / Accepted: date}

\maketitle

\begin{abstract}
The past decade has seen tremendous growth in the field of Complex Social Networks. Several network generation models have been extensively studied to develop an understanding of how real world networks
evolve over time. Two important applications of these models are to study the evolution dynamics and processes that shape a network, and to generate benchmark networks with known community structures. Research has been conducted in both these directions, relatively independent of the other. This creates a disjunct between
real world networks and the networks generated as benchmarks to study community detection algorithms.

In this paper, we propose to study both these application areas together. We introduce a network generation model which is based on evolution dynamics of real world networks and, it can generate networks with community structures that can be used as benchmark graphs. We study the behaviour of different community detection algorithms based on the proposed model and compare it with other models to generate benchmark graphs. Results suggest that the proposed model can generate networks which are not only structurally similar to real world networks but can be used to generate networks with varying community sizes and topologies.

\end{abstract}

\section{Introduction}



The field of complex networks became popular from the late 1990's after the ground breaking discoveries of structural characteristics of small world\cite{watts98} and scale free networks\cite{barabasi99}. Since then, researchers have actively pursued the development of network generation models to mimic the creation and evolution of complex networks emerging from a variety of real world interconnected systems. These models also provide an understanding of other structural characteristics such as assortative mixing\cite{newman02}, presence of hierarchical structures\cite{gilbert11}, presence of communities\cite{ferrara12} and likelihood of connection between similar objects. There is substantial interest in using these synthetic networks to examine the impact of different dynamic processes on these networks like epidemic spreading\cite{pastor01,liang13} and influence mining\cite{kempe03,banos13}.
	
Two important application areas of these models are 1) Study of evolution dynamics and 2) Generating benchmark networks to study community detection algorithms. Evolution dynamics are important because they help us understand how real world networks attain certain structural properties. For example, triadic closures explain the presence of high clustering coefficients, and preferential attachment explains why degree distributions follow power-law. The second application area is in the study of community detection algorithms where these models can help us determine the quality of a community detection algorithm.

Real-world networks emerging from different domains have different topological and structural features. For example, a co-authorship network of collaboration inducts cliques representing a set of authors strongly connected to each other whereas an online social network introduces nodes gradually as new members sign up. This suggests that the underlying mechanism through which these networks evolve,(or simply the evolution dynamics) shape the structural and topological properties of a network. Figure \ref{fig::astr-air} shows the results of studying the building blocks of two networks from different domains. The authors \cite{zaidi10a} diagrammatically show the differences in topological features between a co-authorship network \cite{newman01} and an air transport network \cite{rozenblat08}.

Thus we argue that models to generate benchmark graphs should consider well known evolution dynamics as part of their generation process. Since the performance of community detection algorithms vary with networks of different topological features, having a tunable model will also help us generate networks with desired structural features and thus can be used to evaluate the performance of a community detection algorithm on a wide variety of structurally different networks.





One of the earlier works to generate synthetic networks with ground truth communities is commonly known as GN benchmark \cite{girvan02}. Lancichinetti and Fortunato \cite{lancichinetti09} identified a number of drawbacks in GN benchmark and presented a model to generate synthetic networks of different sizes with desired degree distributions and clustering. The model also provides a mechanism to generate networks with sparse or dense communities. However, it does not consider any dynamics or microscopic rules such as preferential attachment, triadic closure which are shown to be the corner stone of real world evolving networks. A number of models are available in the literature to generate synthetic networks with community structures but these models do not provide the mechanism to generate ground truth communities and ultimately cannot be used to produce benchmark graphs.




This paper is organized as follows: In Section \ref{sec::related_work}, literature for generating benchmark graphs along with models to generate networks with communities are discussed. Section \ref{sec::proposed_model} describes the proposed model whereas section \ref{sec:exprsetup} explains the experimental setup. Section \ref{sec:result} presents the results and we conclude in section \ref{sec::conclusion}.

\section{Related Work}\label{sec::related_work}

\subsection{Benchmark Graphs}

There are two approaches to evaluate the performance of community detection algorithms. First is to test against real-world networks with prior information about communities and the second approach is to test against synthetically generated networks with ground truth communities \cite{newman12}. Community detection algorithms using the first approach have to rely on small networks such as Zachary's karate club \cite{zachary77} due to unavailability of ground truth for large networks. Yang and Leskovec \cite{yang15} studied a few large networks and identified their ground truth communities based on their nodal attributes. Real-world networks not only behave inconsistently for different definitions of communities, but this is also true when nodal attributes are used to cluster nodes as compared to structural characteristics. Hric \textit{et al.} \cite{hric14} found significant differences between ground truth communities and the communities identified by community detection algorithms in real-world networks. 

The second approach to evaluate algorithms is to test against synthetic networks generated by models. Ideally such models must be tunable and capable of generating ground truth communities alongside the network itself, in order to compare the results produced by community detection algorithms. Although, there are a number of models available in the literature to produce synthetic networks with community structures their generation processes and mechanisms make it difficult to generate and/or extract ground truth communities.

One of the earliest work in this direction was introduced by Girvan and Newman which is known as the GN benchmark or 'four-groups' networks \cite{girvan02}. The model generates a network with 128 nodes divided into four groups i.e.\ 32 nodes in each group where the average degree of each node is 16. This is close to a random graph of a similar configuration. In contrast to random graphs, the inter-cluster and intra-cluster connectivity of each node can be controlled by a parameter. Danon \textit{et al.} proposed a variant of GN network to introduce communities of different sizes but failed to mimic real-world networks with communities due to their small size and absence of fat tails in their degree distributions. Lancichinetti proposed another model to generate benchmark networks, known as the LFR benchmark, that can generate networks with heterogeneous degree distributions following power law. This model is capable of generating large size networks with different size communities. It is also capable of generating networks with different topological properties such as degree distributions and average degrees \cite{lancichinetti08}.  The model focusses on generating networks with desired properties but does not use any known evolution dynamics or microscopic rules for the generation process. Furthermore the model uses a minimum and maximum number between which it determines the number of communities. This indeterminism leads to a different number of communities every time the algorithm executes. These reasons motivated us to propose a new model which incorporates well studied microscopic rules for complex networks and essentially use them to generate synthetic networks with communities.


\subsection{Network Models with Community Structure} 

Since the discovery of the ground breaking models to generate small-world \cite{watts98} and scale-free \cite{barabasi99} networks, a number of network models have been proposed which are either variants or extensions of these two models, or use a completely different approach to generate networks. Early models generated networks exhibiting properties of small-world and/or scale-free networks with low average path lengths, high clustering coefficients and degree distributions following power law. With advancements over time, other network structural properties have also gained importance resulting in newer models being proposed. Below we discuss several models to generate networks with community structures.



Xu \textit{et al.} \cite{xu09} extended the scale-free model (BA) \cite{barabasi99} by introducing global random attachment for community selection whereas preferential attachment is used for the selection of nodes similarly as in the BA model. The basic idea is to first select a community randomly from the existing communities in the network followed by the selection of a node on the basis of preferential attachment. The model also creates inter-community and intra-community edges with some probabilities. The authors are able to generate networks with general power law distribution of nodes' degree with the presence of community structures in the network. 

Xie \textit{et al.} \cite{xie07} proposed an evolving model based on preferential mechanism for the selection of communities and nodes to generate networks with community structures. Communities of larger sizes are preferred at the time of selection of a community, and a high degree node is preferred to select neighbours to create inter-cluster edges. A similar model was proposed by Zhou \textit{et al.}  \cite{zhou08} for weighted evolving networks with additional triad formation step to achieve high clustering in generated networks. Both models allow the introduction of a fully connected community or a single node into the network with some probability.

The model by \cite{kumpula09} is based on microscopic rules from sociology to generate networks with moderate size communities. They used cyclic closure and focal closure which are proven mechanisms in sociology to acquire new acquaintances in social networks. Random walk is used to achieve cyclic closures in order to find strong ties in the neighbourhood of a node whereas focal closure is achieved by formation of links with random nodes in the network.

Zaidi \cite{zaidi13a} discussed the role of randomness in the generation of community structures in a network. They introduced different sizes of closely connected communities by replacing nodes in a network generated by the ER model \cite{erdos60}. The authors found that by introducing a small order in random networks, we can generate clustered networks. However, networks generated by this approach do not have the scale-free property which is observed in many real world networks. 

The model proposed by Sallaberry \textit{et al.} is similar to configuration models such as Generalized Random Graph Models as it generates a network for the given degree distribution. However, the model generates cliques for the given degree distribution and creates connections among these cliques on the basis of distances among them \cite{zaidi13b}. The model is static in the sense that the number of nodes remain constant once degree distribution is assigned to initial nodes. 

Recently an extension of \cite{holme02} was proposed by Pasta \textit{et al.} \cite{pasta13a} which is based on global random selection of a community and local preferential attachment for the selection of a node. The model produced networks with three important structural properties that are high clustering coefficient, presence of hierarchical community structure, and each community following power law distribution for nodes' degree.


These models use different microscopic rules to generate networks with different topological structures but do not preserve information about the community memberships, hence cannot be used as benchmarks.



\section{Proposed Model}\label{sec::proposed_model}                                                                                                                                                                              

The proposed model uses two microscopic rules, global community selection and local preferential attachment to generate a network with community structures. The model comprises of similar steps as the original model proposed by \cite{pasta13a} with additional heuristics to decide over the community of each node. A mixing parameter ($\mu$) which ranges between 0 and 1 is introduced. This parameter determines the quality of modular structures in the network. Lower values of $\mu$ result in well separated communities and vice versa. Each new node has $\mu$ fraction of edges with nodes belonging to the same community whereas $1 - \mu$ edges with nodes belonging to other communities. Other controlling parameters include: number of nodes in network ($N$), number of communities in network ($\varsigma$), probability of triad formation($P_{t}$), number of edges for each new node($m$). The model has two major phases:

\textbf{Setup:} Initially we introduce as many triads as the required number of communities ($\varsigma$). Each triad represents a community and each node of the respective triad is labelled with a number representing its community. 

\textbf{Growth:} The following steps are repeated until number of nodes reaches $N$ 

\begin{enumerate}
\item
At each timestamp $t$ a new node $n$ is introduced to the community and connected to an existing node $n^{'}$ selected on the basis of preferential attachment. At this stage the community of node $n$ is assigned to node $n^{'}$ as well. Thus, both nodes now belong to the same community and have similar labels. 


\item
With probability $P_{t}$, node $n$ creates $m$ edges with neighbours of $n^{'}$ whereas fraction $1-\mu$ edges are connected to neighbours which are not part of the same community and a fraction of $\mu$ edges are created with the neighbours of the same community. 

\item
Node $n$ then creates $1-m$ edges preferentially with probability $1 - P_{t}$ whereas fraction $\mu$ edges connect to nodes from the same community while a fraction $1 - \mu$ edges are connected to nodes of other communities. 
\end{enumerate}


\section{Experimental Setup}
\label{sec:exprsetup}

The proposed model was tested for different configurations of four controlling parameters as described below. The parameter to control number of edges for each new node ($m$) is fixed to 2 for all configurations. We produced 240 graphs and the results were averaged over 5 instances for each of the configuration to reduce biasness caused by randomness. 
\begin{itemize}
\item \textbf{Networks Size $N$} (Networks of 1000, 2000 and 4000 nodes)
\item \textbf{Number of communities $\varsigma$}(10, 20, 30, and 40 communities)
\item \textbf{Probability of Triad Formation $P_{t}$} (0.1, 0.3, 0.5, 0.7 and 0.9)
\item \textbf{Mixing Parameter $\mu$}(0.2, 0.4, 0.6, and 0.8)
\end{itemize}





\subsection{Goodness of Community}
\label{goodness}
Despite the fact that community detection is one of the most widely studied problems in network science, there exists varying quantifiable definitions for a \textit{community}. Thus, testing a benchmark against a specific definition may lead to a situation in which a benchmark does not provide consistent results for all community detection algorithms. Thus the proposed model was tested against different definitions such as separability, density, clustering coefficient and loyalty. 

\textbf{Separability:} defines how well a community is separated from the rest of the network i.e. a good community must have a low number of edges pointing to nodes outside it's community\cite{shi00}. This can be quantified as the ratio between number of edges inside and outside the community. Let $C$ be the community in consideration, V as the set of nodes and E is the set of all edges, we define separability as:
\begin{equation}
{f(C) = \frac{ |\{(u,v) \in E: u \in C, v \in C\} | }{|\{(u,v) \in E: u \in C, v \notin C\} | }}
\end{equation}

\textbf{Density:} quantifies the fraction of edges which are part of the same community from all possible edges \cite{fortunato10}. The more edges within a community, the better the community - regardless of the number of edges pointing to nodes of other communities. Here $n_{c}$ is the number of nodes in community C.
\begin{equation}
{f(C) = \frac{ |\{(u,v) \in E: u \in C, v \in C\} |}{n_{C}(n_{C}-1)/2}}
\end{equation}

\textbf{Clustering Coefficient:} Another definition of community is based on close binding of neighbours in the same community. This can be quantified as clustering coefficient \cite{watts98}. A community of nodes with high clustering coefficient is better than a community with low clustering coefficient. 

\textbf{Loyalty:} of a node is the fraction of edges of a node connected to nodes of the same community. A community with disloyal nodes suggests that the these nodes have more edges to other communities rather than to nodes of their own community. The \textit{loyalty} of a community is the average \textit{loyalty} of all the nodes in it's community. 
\begin{equation}
{f(C) = \frac{ |\{(u,v) \in E: u \in C, v \in C\} | }{|\{(u,v) \in E: u \in C, v \in V\} |}}
\end{equation}
The \textit{Separability} and \textit{Loyalty} both capture the same intuition that a community must be separated from the rest of the network but \textit{Loyalty} quantifies this in a range of 0 and 1. This makes it easier to compare the results of two distinct sets of communities generated as a result of different community detection algorithms. 

\subsection{Community Detection Algorithms}

The presence of community structures is one of the most important characteristics of real world networks. Finding communities in a network is a difficult but crucial task to understand the internal structures of a network. This has been a widely studied area by scholars of different domains and a number of algorithms have been proposed to detect communities in networks. 

In order to cover a wide spectrum of community detection algorithms, eight community detection algorithms were selected for the experiment carried out in this paper and each algorithm represents a distinct class of community detection algorithms. The selected algorithms are: Fast greedy clustering is a modularity optimization algorithm\cite{clauset04},  Multilevel clustering which is an extension of modularity optimization with additional steps to merge found communities into a single node to repeat the same process \cite{blondel08}, Walktrap algorithm is based on random walks to calculate distances between nodes in order to group them in one cluster \cite{pons05}, Markov Clustering (MCL) uses markov matrices to simulate stochastic flow \cite{vandongen00}, Infomap clustering which solves the community detection problem using Information Theory \cite{rosvall08}, Label Propagation which only uses network structure without any prior information about communities or any objective function \cite{raghavan07}, VOS which uses the network layout approach to determine communities \cite{waltman10}, and Spinglass based on the statistical mechanics' spin models \cite{reichardt06}. The detailed review of community detection algorithms is out of scope for this article and readers can refer respective citations for further details or \cite{leskovec10, orman11} for comparative analysis. 

The performance of these eight algorithms on networks generated by LFR benchmark and the proposed model were evaluated. There are a number of criterion available to compare ground truth communities with results generated by community detection algorithm. This paper used the most commonly used measure \textit{Normalized Mutual Information} (NMI), in which the value is equal to 1 when two partitions are identical to each other, whereas it has an expected value of zero when partitions are dissimilar to each other \cite{danon05}.
%

\section{Results and Discussion}
\label{sec:result}

\subsection{Evaluation of Structural Properties}

First we evaluate the structural properties of the networks generated using the proposed model. The three main properties studied here are the Degree Distribution, the Average Path Length and the Clustering Coefficient.

One of the fundamental features of complex networks is that their degree distribution follows power law. Figure \ref{fig::str_degree} shows that for different values of the mixing parameter $\mu$, the generated graphs indeed have a fat tail distribution with power law coefficient $\alpha$ ranging between $[2,3]$.


Figure \ref{fig::n_aplcc} shows the behavior of Average Path Length (APL) and Clustering coefficient (CC) with respect to Network Size ($N$) and Probability of Triage Formation ($P_t$). Figure \ref{fig::str_apl} shows that APL increases with increasing network size $N$ as well as increasing probability of Triad formation ($P_t$). As more nodes are introduced in the network, the average geodesic distance among the nodes increases. Similarly, as more triads are formed, more short range edges are created, increasing the overall average distance among the nodes. Figure \ref{fig::str_cc} shows that CC increases as $P_{t}$ increases irrespective of the the network size $N$. This is obvious as a higher probability of triad formation introduces more triadic closures increasing the overall clustering coefficient of the network. Only a negligible variation in clustering coefficient is observed for different values of $N$. 


Figure \ref{fig::mu_aplcc} studies the behaviour of Mixing Parameter $\mu$ on Average Path Length (APL) and Clustering Coefficient (CC). Figure \ref{fig::fig_apl} shows that for lower values of $\mu$, the communities are more separated, thus resulting in nodes of a community to be closer to each other but further apart from nodes from other communities. This results in lower overall APL for the network. As the number of communities is higher ($\varsigma=40$), the APL is higher because the nodes are segregated across communities when compared to lower number of communities ($\varsigma=20$). Impact of size of network can also be observed as more nodes again result in higher APL for the generated networks. Figure \ref{fig::fig_cc} shows that higher values of $\mu$ result in slightly lower values of clustering coefficient. This is simply because an increased number of nodes and edges across communities end up not being part of triads resulting in lower overall clustering coefficient. Irrespective of the size of networks, higher $P_t$ values result in higher clustering coefficients as expected.


\subsection{Presence of Communities}

In this subsection, we study the behaviour of the communities generated by the proposed model. First, we use Modularity \cite{newman04} to quantify the presence and quality of ground truth community structures in the  generated networks. The values of modularity for different graphs vary as a function of mixing parameter $\mu$. Figure \ref{fig::m1} shows that the quality of modular structures decreases for increasing values of mixing parameter and the behaviour is consistent for various number of communities generated in the networks.

Figure \ref{fig::fig_g1} shows the values of four goodness of community metrics for different values of $\mu$. Y-axis shows that some communities better fit to the respective definition of community as compared to others. However, it is visible that as the value of mixing parameter increases the more communities poorly fit with values approaching to $0$. This represents that communities in the network generally do not fit well for the respective definition. Similar results were observed for different sizes of network which indicates that structurally different communities can be generated in a network by tuning the values of mixing parameter.




The performance of eight clustering algorithms were compared on networks generated using the proposed model and the LFR model. Normalized Mutual Information (NMI) was calculated to quantify the similarity between generated communities and the ground truth communities. Figure \ref{fig::comp_avg} shows the average for all instances of respective benchmarks for different sizes. Community detection algorithms performed poorly for  higher values of $\mu$ as expected. However, it is clearly visible that graphs generated by LFR benchmarks are more sensitive than graphs generated by the proposed model for mixing parameter. It indicates that overall community detection algorithms do not behave similarly for these two models to generate benchmark networks. 

Figure \ref{fig::compind} studies individual performances of the community detection algorithms. Each graph shows the result of one algorithm for LFR benchmark (dotted line) and the proposed model (solid line). The algorithms do not behave similarly for these two benchmarks. Generally a more consistent behaviour can be observed for the networks generated using the proposed model. All community detection algorithms have a similar trend with NMI values inversely proportional to mixing parameter $/mu$. For the LFR Benchmark there are some inconsistencies in the generated networks as VOS and MCL clustering algorithms stand out as better performers in detecting the ground truth communities even with the increasing $/mu$ values. Apart from these two clustering algorithms, the networks generated through LFR show higher sensitivity to the mixing parameter as compared to the proposed model where higher $/mu$ values result in finding poor communities. These observations simply highlight the fact that irrespective of the similarity in structural properties such as clustering coefficient and average path length, the topology of the network changes with the methodology used to generate networks. This in effect, changes the structure of communities in turn resulting in topologically different networks. Note that we are not trying to prove that the proposed model is better than the LFR model, but simply to the fact that the networks and communities generated by these two models are topologically different and community detection algorithms should consider these differences when studying simulated benchmark networks.



\subsection{Comparison with Real World Networks}

Using the proposed model, we also generated networks similar to three topologically different real world networks. This is to show that the proposed model is quite flexible and can be used to produce networks equivalent to other networks with desired structural properties. The proposed model successfully generated networks which are structurally similar to real world networks when compared using average path length, clustering coefficient and exponent of the fitted power-law distribution. We also used Holme and Kim model \cite{holme02} to generate networks similar to real world networks.

\begin{table}
\centering
\label{table:parameters-hk}
\begin{tabular}{|l|l|l|}
\hline
\multicolumn{3}{|l|}{\textbf{Parameters used to generate networks using Holme \& Kim Model}} \\ \hline
\textbf{Dataset}                       & \textbf{Pt}             & \textbf{m}             \\ \hline
Air Transport Network                  & 0.9                    &  12\\ \hline
Network of Political Books             & 0.9                    & 5                  \\ \hline
Co-author Geometry                     & 0.9                    & 3                  \\ \hline
\end{tabular}
\caption{These parameters were used to generate networks similar to real-world networks using Holme \& Kim model \cite{holme02}. $P_{t}$ = Probability of Triad Formation, $m$ = number of edges for each new node}
\end{table}

\begin{table}
\centering
\label{table:parameters-benchmark}
\begin{tabular}{|l|c|c|c|c|}
\hline
\multicolumn{5}{|l|}{\textbf{Parameters used to generate networks using the proposed model}} \\ \hline
\textbf{Dataset}             & \textbf{$N_{o}$}   & \textbf{$P_{t}$}   & \textbf{$\mu$}  & \textbf{$m$}  \\ \hline
Air Transport Network        & 30            & 0.9           & 0.3          & 12          \\ \hline
Political Books Network      & 3             & 0.9           & 0.2          & 5           \\ \hline
Co-author Geometry Network           & 40            & 0.9           & 0.3          & 3           \\ \hline
\end{tabular}
\caption{These parameters were used to generate networks similar to real-world networks using the proposed model. $N_{o}$= number of communities in the network,  $P_{t}$ = Probability of Triad Formation, $\mu$ = mixing parameter, $m$ = number of edges for each new node}
\end{table}

\begin{table*}
\centering
\label{tbl:realnetworks}
\begin{tabular}{|l|c|c|c|c|c|c|}
\hline
\multicolumn{7}{|c|}{\textbf{Comparing Real World Networks and Networks Generated }} \\ 
\multicolumn{7}{|c|}{\textbf{using Holme \& Kim, and the Proposed Model}} \\ \hline
                       & Nodes  & Edges    & Node/Edge     & APL     & CC      & $\alpha$   \\ \hline
Air Transport Network       & 1540   & 16524    & 10.72         & 4.24    & 0.26    & 2.68     \\ \hline
Holme \& Kim                & 1540   & 18447    & 11.97         & 2.59    & 0.11    & 2.86     \\ \hline
Proposed Model              & 1540   & 17226    & 11.18         & 3.98    & .021    & 2.86     \\ \hline
\hline
Political Books Network             & 105    & 441      & 4.2   & 3.0     & 0.34    & 2.62   \\ \hline
Holme \& Kim                & 105    & 513      & 4.88          & 2.29    & 0.28    & 2.71     \\ \hline
Proposed Model              & 105    & 486      & 4.62          & 2.77    & 0.35    & 2.62     \\ \hline
\hline
Co-Author Geometry Network  & 3621   & 9461     & 2.6           & 5.31    & 0.22    & 2.45     \\ \hline
Holme \& Kim                & 3621   & 10857    & 2.9           & 4.51    & 0.14    & 2.72     \\ \hline
Proposed Model              & 3621   & 10663    & 2.9           & 5.98    & 0.21    & 2.68     \\ \hline
\end{tabular}
\caption{Comparison of Three Real World Networks: Air Transport Network \cite{rozenblat08}, Network of Political Books \cite{polbooks} and Co-Author Geometry Network \cite{geomdataset} compared with the model by Holme \& Kim \cite{holme02} and the proposed model. APL: Average Path Length, CC: Global Clustering Coefficient, $\alpha$: exponent of the fitted power-law distribution}
\end{table*}

\section{Conclusion}\label{sec::conclusion}
In this paper, we proposed a new model to generate benchmark graphs with communities. The model is based on evolution dynamics and microscopic rules such as preferential attachment and triadic closure. The proposed model can not only  generate scale-free and small-world networks with communities, but is also flexible enough to generate networks similar to real world networks. The performance of eight different community detection algorithms was studied and compared with the state of the art LFR benchmark. The various community detection algorithms demonstrated inconsistent behaviour on the two benchmarks, the newly proposed model and the LFR highlighting the fact that different microscopic rules indeed affect the topology of generated networks. Thus proving our initial hypothesis that benchmark networks should incorporate well known evolution dynamics in order to study community detection algorithms.


As part of our future work, we intend to investigate reasons behind the behaviour of community detection algorithms on both benchmarks. There are numerous microscopic rules available in the literature. We also aim to study the behaviour of community detection algorithms against different microscopic rules available in literature. Another area that we want to explore is the study of the quality of communities generated by benchmark networks.

\section{Declarations}
\subsection{List of abbreviations}
This list shows the abbreviations in the alphabetical order:

\begin{itemize}

\item \textit{CC:} Clustering Coefficient  
\item \textit{APL:} Average Path Length 
\item \textit{MCL:} Markov Clustering 
\item \textit{NMI:} Normalized Mutual Information

\end{itemize}

\subsection{Availability of data and materials}
The datasets supporting the conclusions of this article are available online and can be accessible at https://dx.doi.org/10.6084/m9.figshare.4560427.

\subsection{Competing Interest}
The authors declare that they have no competing interests.

\subsection{Authors' contribution}
Authors contributed to the manuscript with the order they appear. All authors discussed the experiments and the final results as well as read and approved the final manuscript.

\bibliographystyle{plain}
\bibliography{visu}

\section{Legends}
\begin{figure}[h!]
\center
\textbf{Building blocks of two networks from different domains}
\includegraphics[scale=0.4]{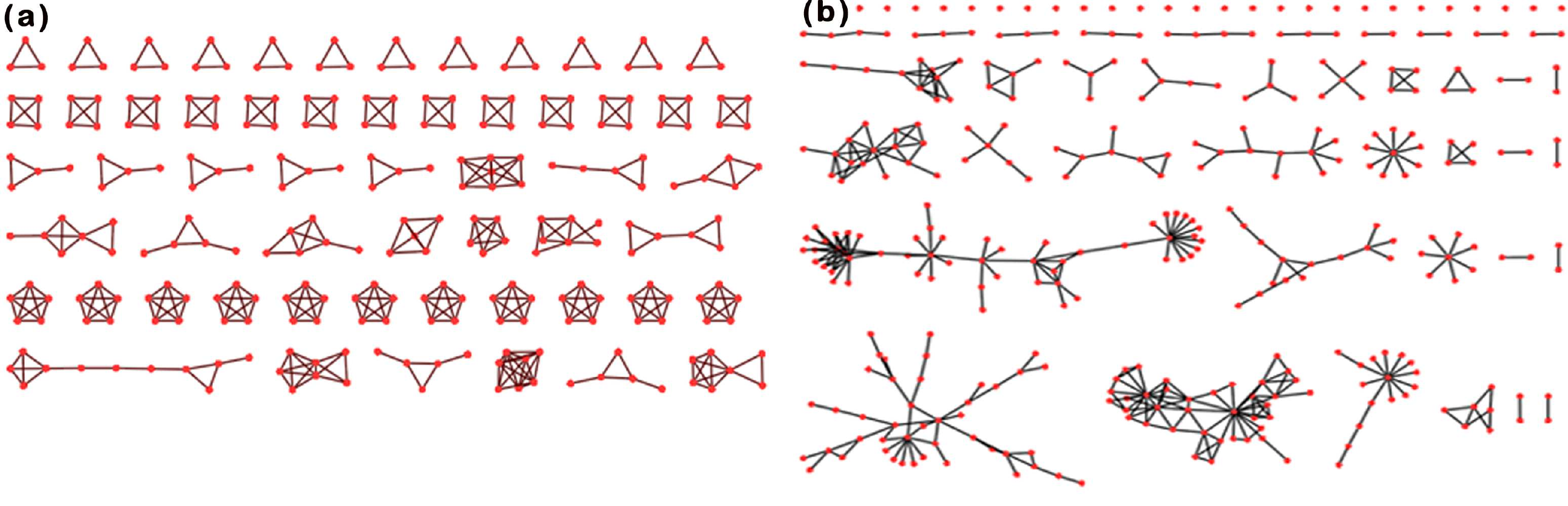}
\caption{Building blocks of two different networks extracted at 5\% of their maximum degree using \cite{zaidi10a}. a) Collaboration network of astrophysics archives \cite{newman01} and b) Air transport network of Cities \cite{rozenblat08}. Clearly the building components of both networks are structurally different as one contains cliques and the other doesn't.}
\label{fig::astr-air}
\end{figure}

\textbf{Figure \ref{fig::astr-air}:} 
\begin{enumerate}
\item Short Title: Building blocks of two networks from different domains
\item Detail Legend: Building blocks of two different networks extracted at 5\% of their maximum degree using \cite{zaidi10a}. a) Collaboration network of astrophysics archives \cite{newman01} and b) Air transport network of Cities \cite{rozenblat08}. Clearly the building components of both networks are structurally different as one contains cliques and the other doesn't.
\end{enumerate}

\begin{figure}
\center
\includegraphics[trim={0cm .3cm .8cm 2cm},clip, width=0.45\textwidth]{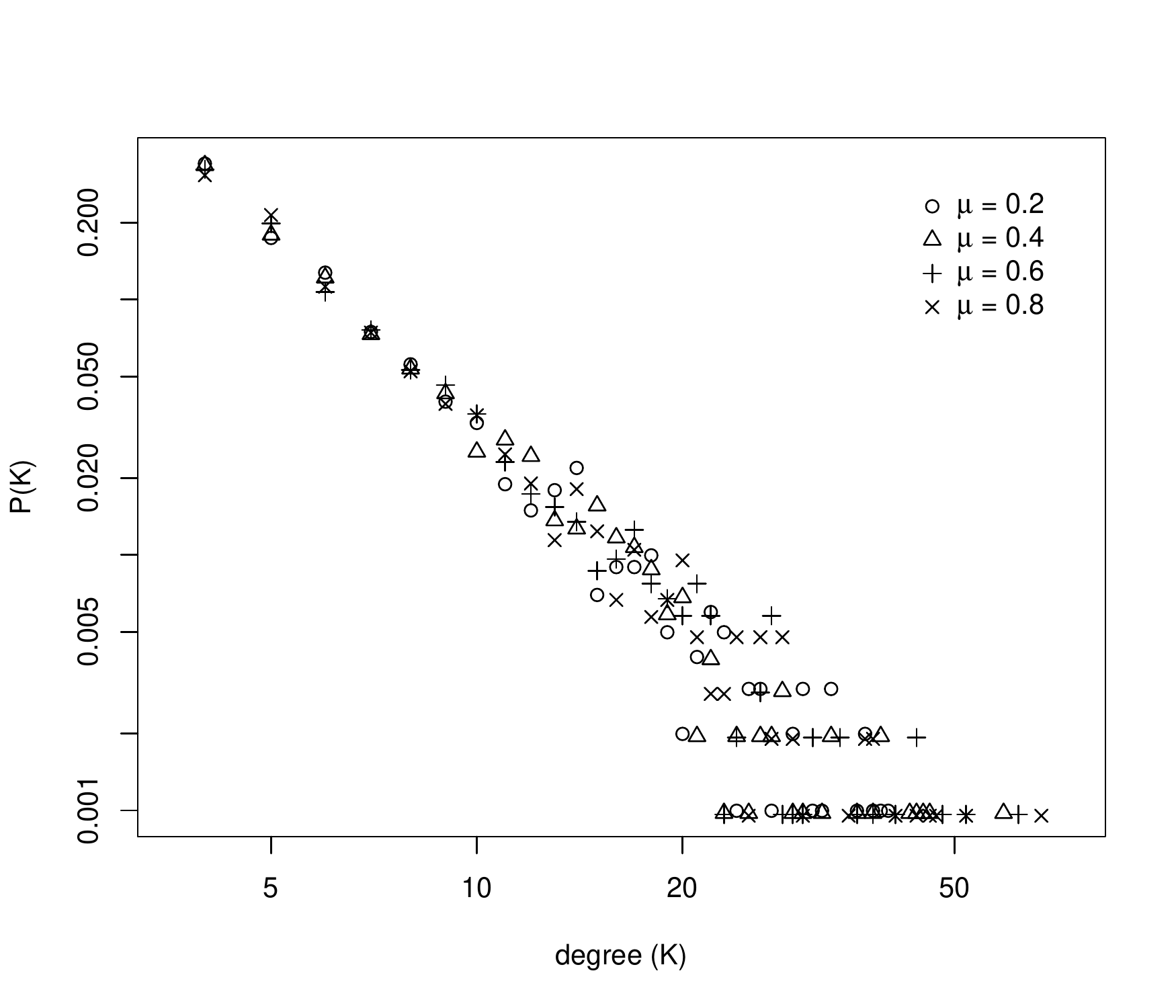}
\caption{Degree distribution of generated networks with different values of mixing parameter ($\mu$)}
\setlength{\belowcaptionskip}{-10pt}
\label{fig::str_degree}
\end{figure}

\textbf{Figure \ref{fig::str_degree}:}
\begin{enumerate}
\item Short Title: Degree Distribution
\item Detail Legend: Degree distribution of generated networks with different values of mixing parameter ($\mu$)
\end{enumerate}

\begin{figure}
\center
\begin{subfigure}{0.45\linewidth}
\includegraphics[trim={0cm 0cm 0cm 0.8cm},clip,width=\textwidth]{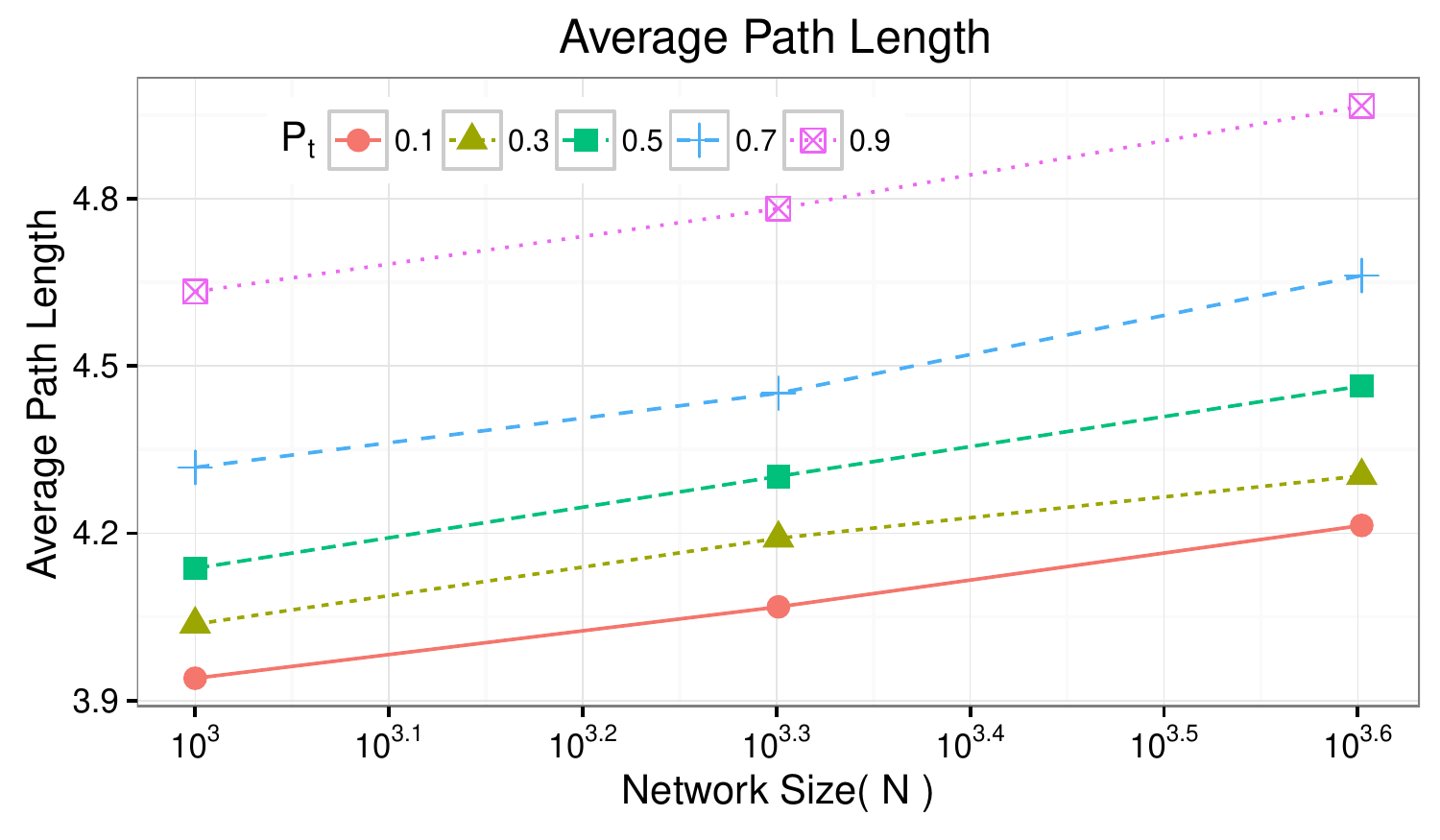}
\caption{Impact of network size ($N$) and Probability of Triad Formation ($P_{t}$) on Average path length of network}
\setlength{\belowcaptionskip}{-10pt}
\label{fig::str_apl}
\end{subfigure}
~
\begin{subfigure}{0.45\linewidth}
\includegraphics[trim={0cm 0cm 0cm .8cm},clip,width=\textwidth]{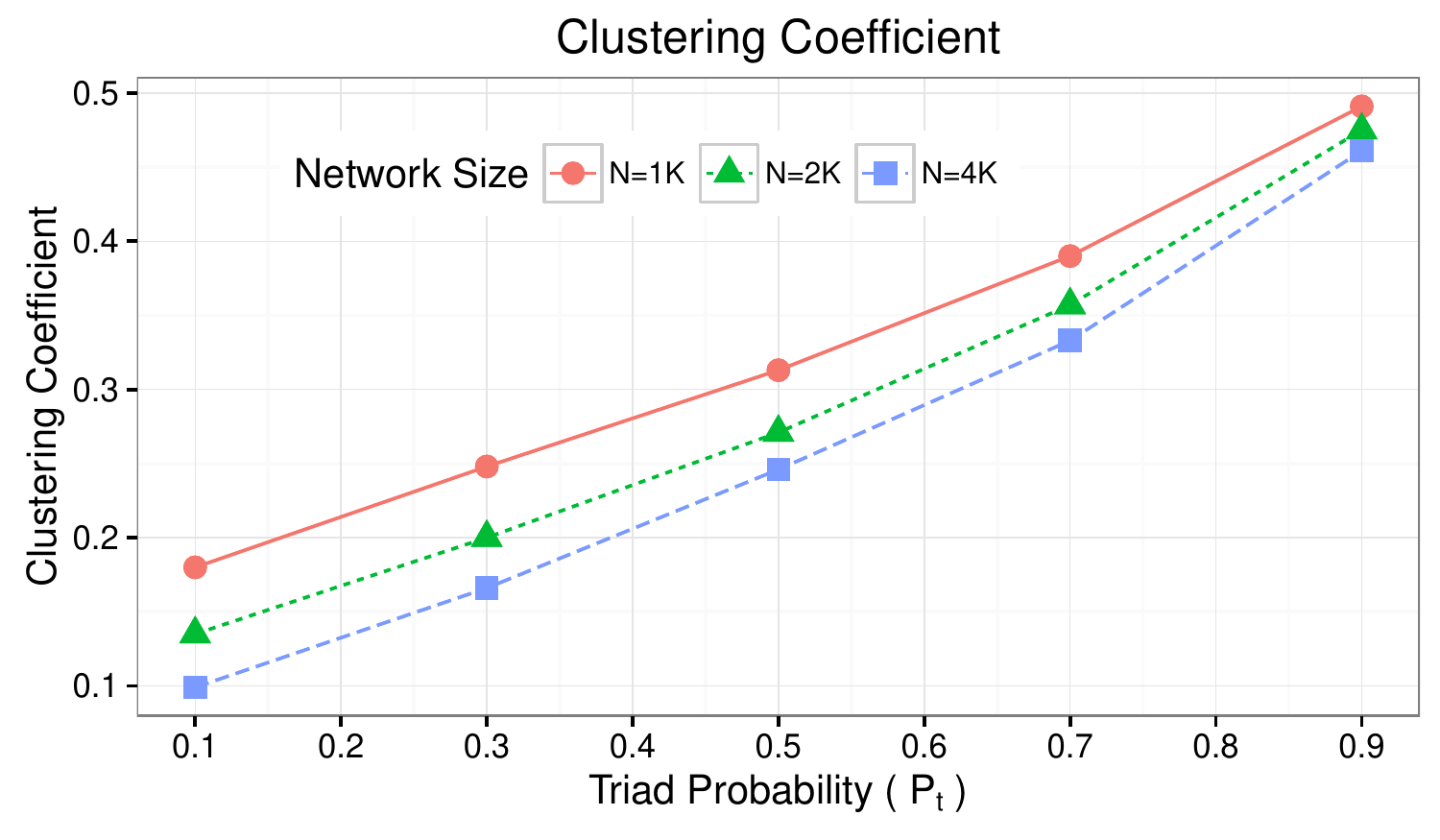}
\caption{Impact of network size ($N$) and Probability of Triad Formation ($P_{t}$) on Clustering coefficient of network}
\label{fig::str_cc}
\end{subfigure}
\caption{Behaviour of Average Path Length and Clustering Coefficient with respect to Network Size and Probability of triad Formation.}
\label{fig::n_aplcc}
\end{figure}

\textbf{Figure \ref{fig::n_aplcc}:}
\begin{enumerate}
\item Short Title: Structural Properties
\item Detailed Legend: Behaviour of Average Path Length and Clustering Coefficient with respect to Network Size and Probability of triad Formation.
\end{enumerate}

\begin{figure}
\center
\begin{subfigure}{0.45\linewidth}
\centering
\includegraphics[width=\textwidth]{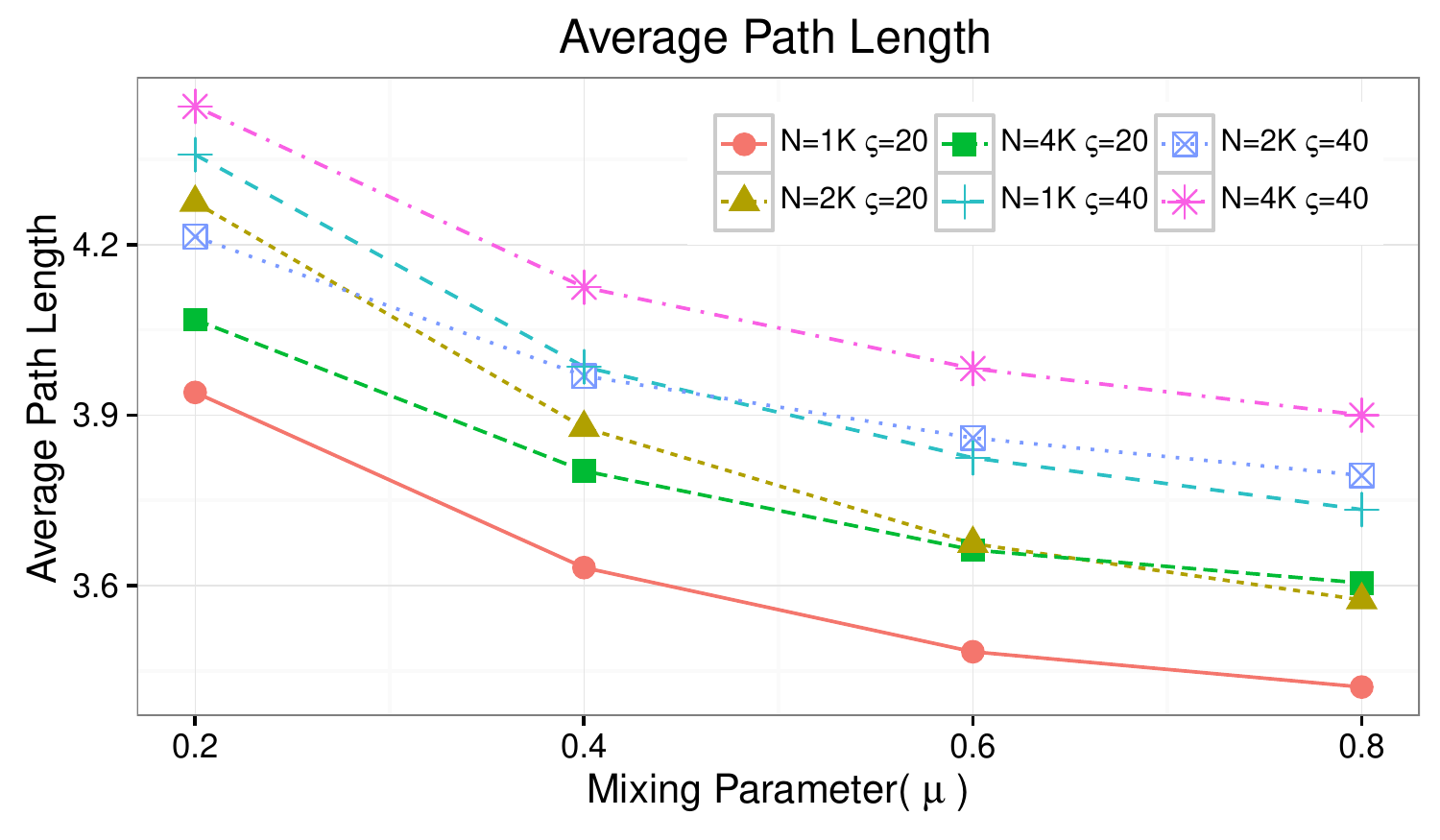}
\caption{N=$\{1000,2000,4000\}$ \& ($\varsigma$)= $\{20,40\}$}
\label{fig::fig_apl}
\end{subfigure}
~
\begin{subfigure}{0.45\linewidth}
\includegraphics[width=\textwidth]{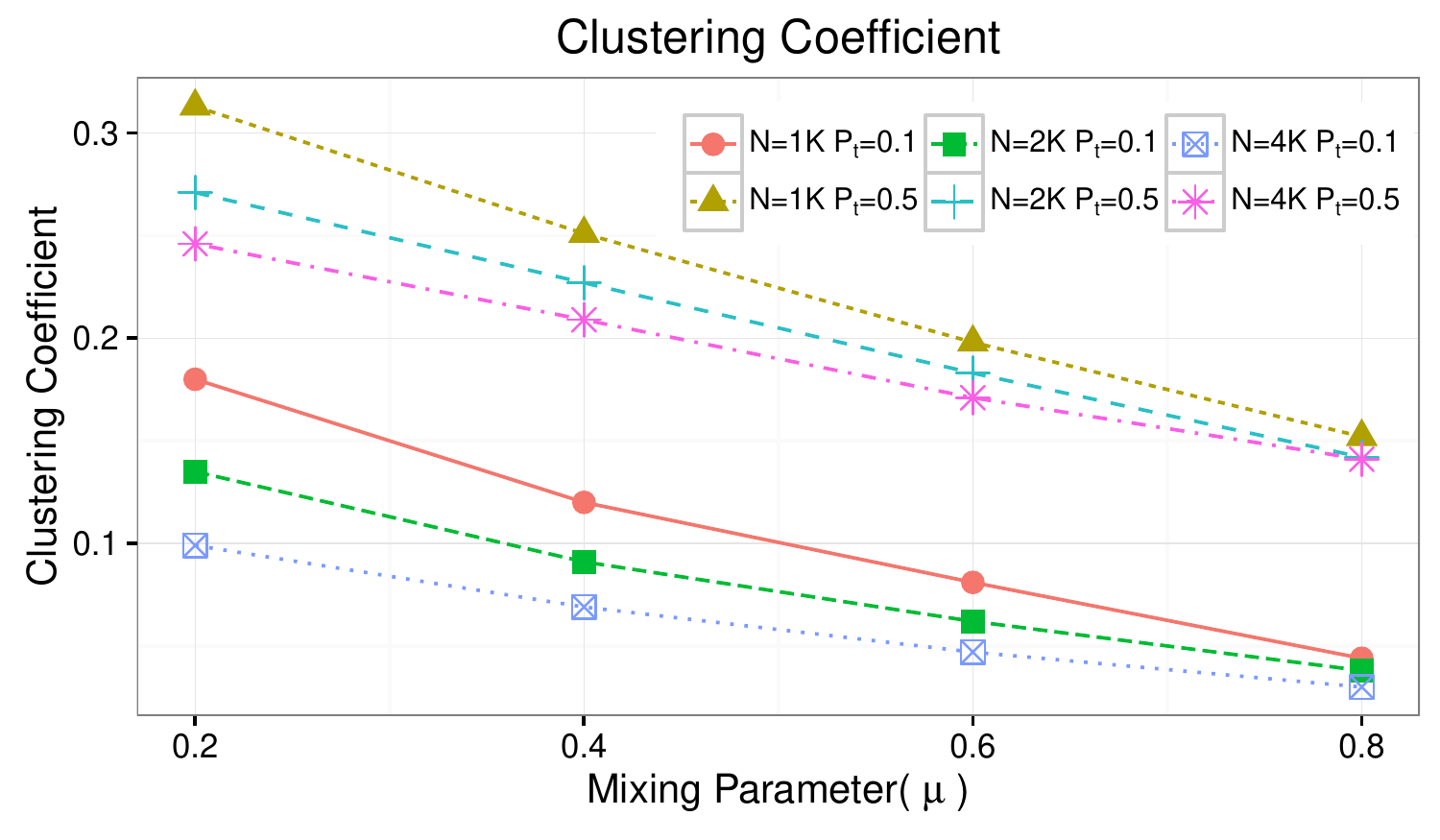}
\caption{N=$\{1000,2000,4000\}$ \& $P_{t}$ = $\{0.1,0.5\}$}
\label{fig::fig_cc}
\end{subfigure}
\caption{Impact of mixing parameter ($\mu$) on (a) average path length (b) overall clustering coefficient}
\label{fig::mu_aplcc}
\end{figure}

\textbf{Figure \ref{fig::mu_aplcc}:}
\begin{enumerate}

\item Short Title: Impact of Mixing Parameter
\item Detailed Legend: Impact of mixing parameter ($\mu$) on (a) average path length (b) overall clustering coefficient
\end{enumerate}

\begin{figure}
\begin{center}
\includegraphics[width=.8\textwidth]{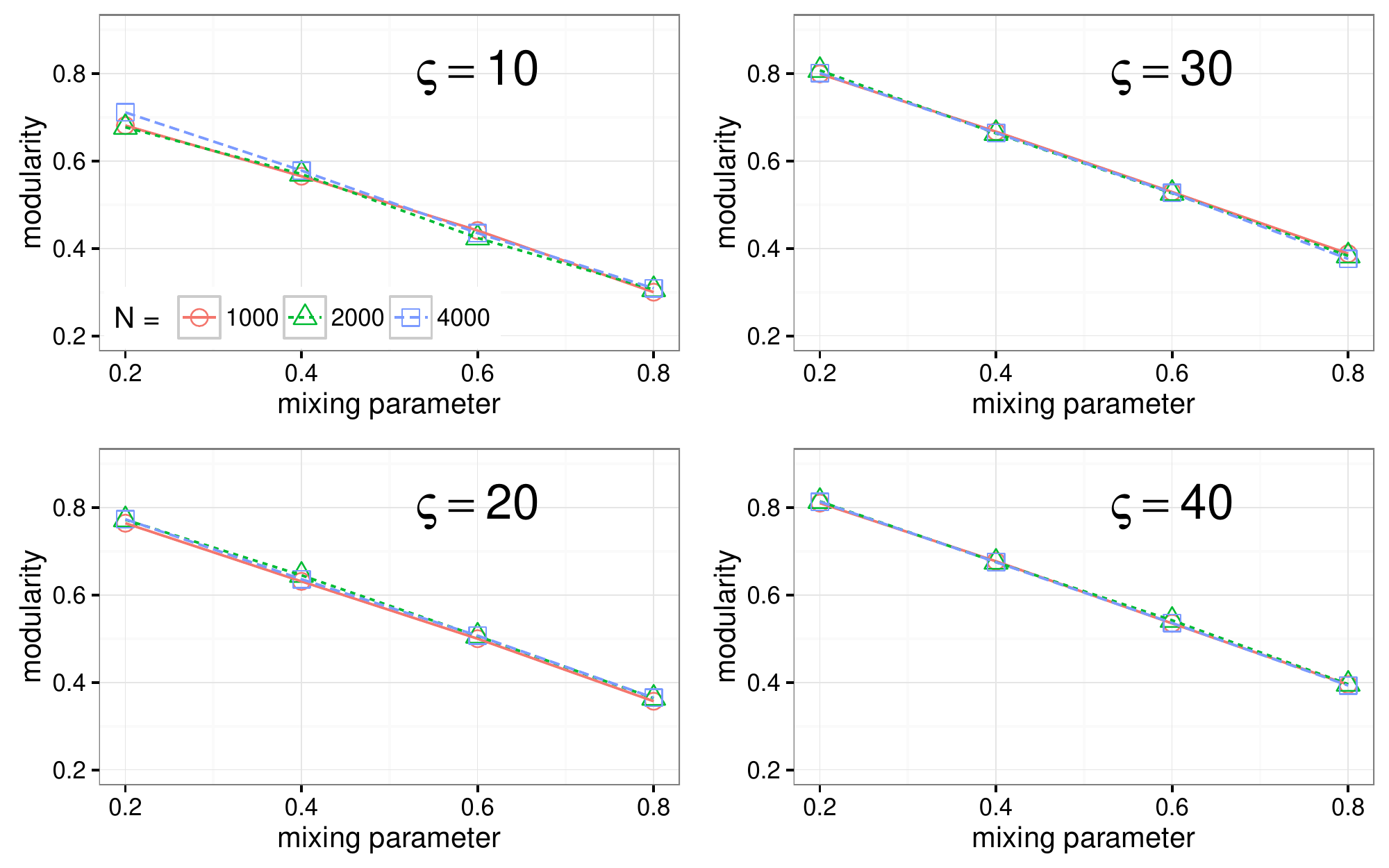}
\end{center}
\caption{\textbf{Test of modularity.} Modularity is indirectly proportional to $\mu$ and the behaviour is consistent for different values of network size $N$ and size of communities ($\varsigma$).} 

\label{fig::m1}
\end{figure}

\textbf{Figure \ref{fig::m1}:}
\begin{enumerate}
\item Short Title: Test of Modularity
\item Detailed Legend:  Modularity is indirectly proportional to $\mu$ and the behaviour is consistent for different values of network size $N$ and size of communities ($\varsigma$).
\end{enumerate}

\begin{figure}
\begin{subfigure}{0.48\textwidth}
\includegraphics[width=\textwidth]{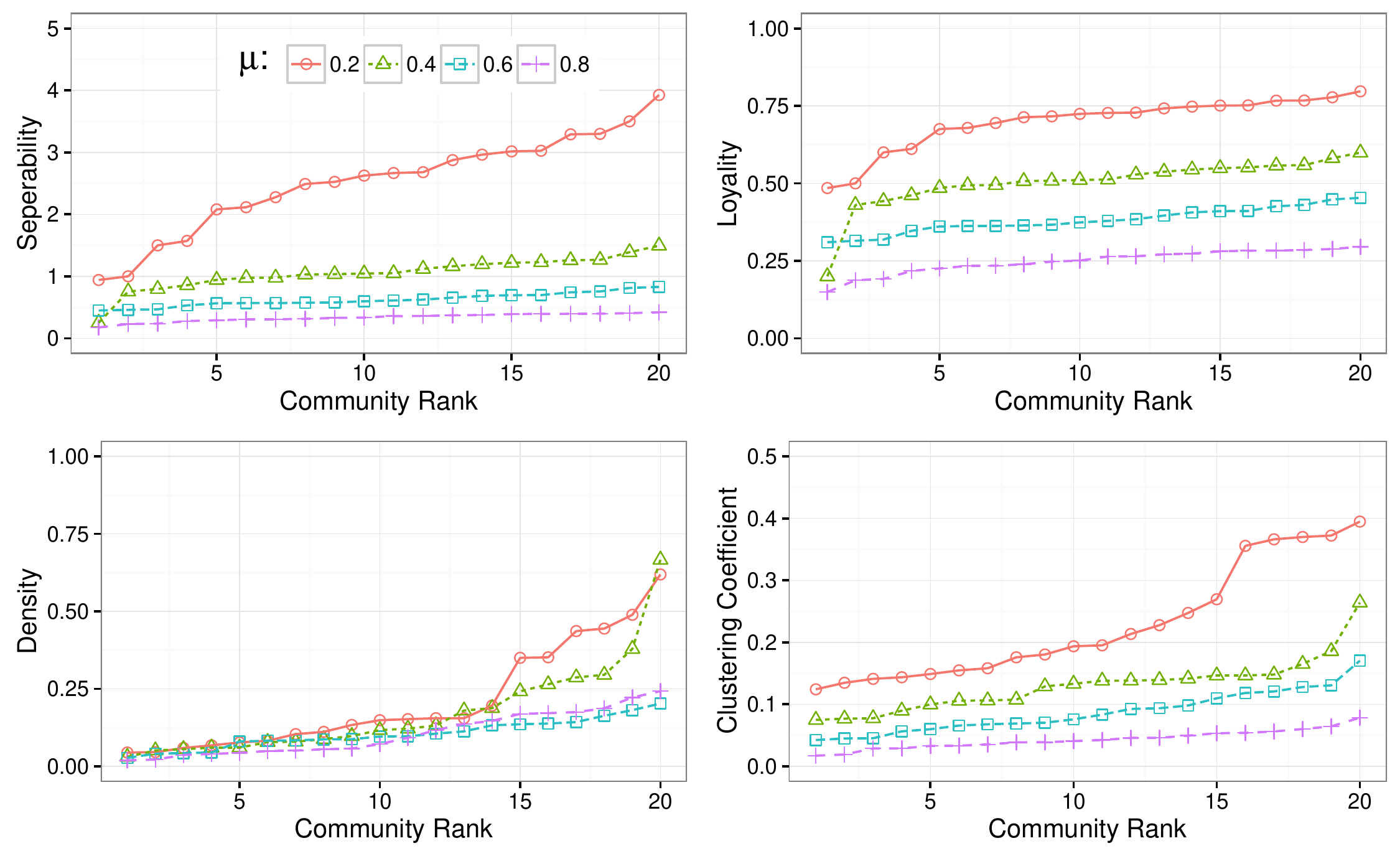}
\caption{Network of size 1000 nodes}
\label{fig::fig_g1.1}
\end{subfigure}
\begin{subfigure}{0.48\textwidth}
\includegraphics[width=\textwidth]{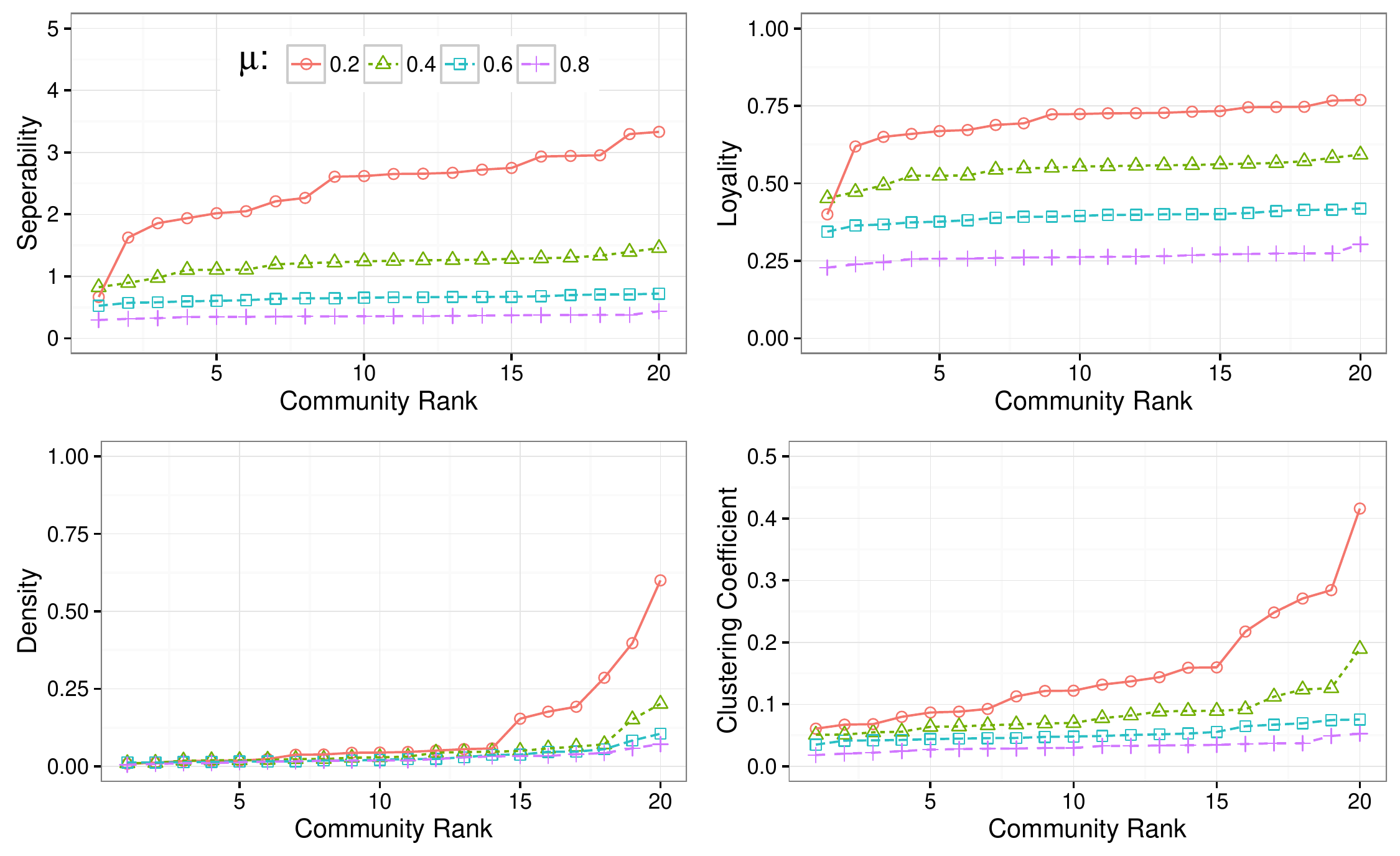}
\caption{Network of size 4000 nodes}
\label{fig::fig_g1.2}
\end{subfigure}
\caption{Goodness of community metrics with triad formation probability ($P_{t}$)=0.1 and Number of communities($\varsigma=20$). Y-axis represents the value of specific goodness metric which represents how good the communities are in the network and x-axis shows the 20 communities labelled 1 to 20. Each line represents the behaviour of communities for a network generated with a specific value of a mixing parameter.}
\label{fig::fig_g1}
\end{figure}

\textbf{Figure \ref{fig::fig_g1}:}
\begin{enumerate}
\item Short Title: Goodness of Community
\item Detailed Legend: Goodness of community metrics with triad formation probability ($P_{t}$)=0.1 and Number of communities($\varsigma=20$). Y-axis represents the value of specific goodness metric which represents how good the communities are in the network and x-axis shows the 20 communities labelled 1 to 20. Each line represents the behaviour of communities for a network generated with a specific value of a mixing parameter.
\end{enumerate}

\begin{figure}
\begin{center}
\includegraphics[width=0.4\textwidth]{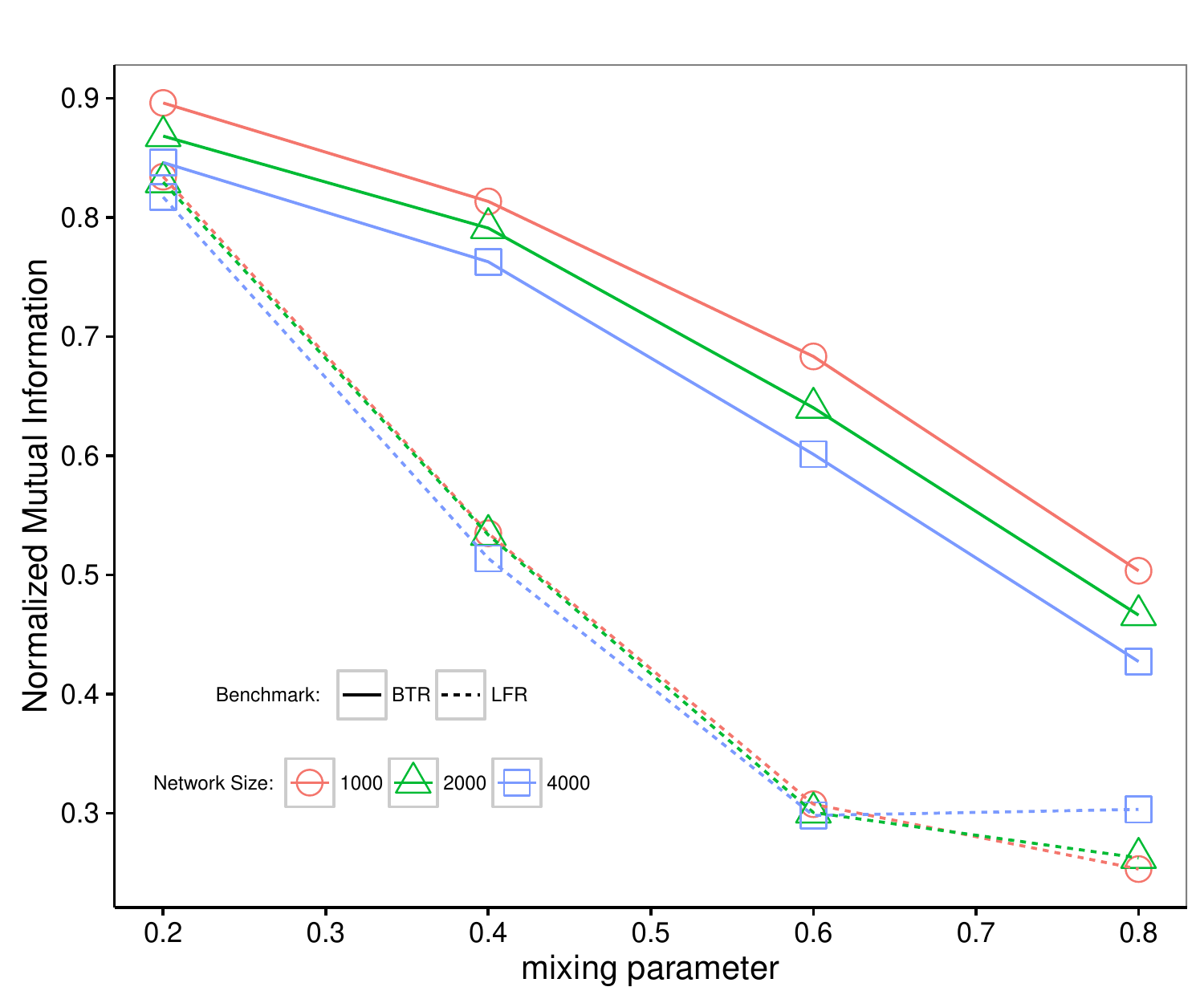}
\caption{Comparison of community detection algorithms}
\label{fig::comp_avg}
\end{center}
\end{figure}

\textbf{Figure \ref{fig::comp_avg}:}
\begin{enumerate}
\item Short Title: Summary of Community Detection Algorithms
\item Detailed Legend: Comparison of eight different community detection algorithms
\end{enumerate}

\begin{figure}
\center
\begin{subfigure}[t]{0.48\textwidth}
\includegraphics[width=\textwidth]{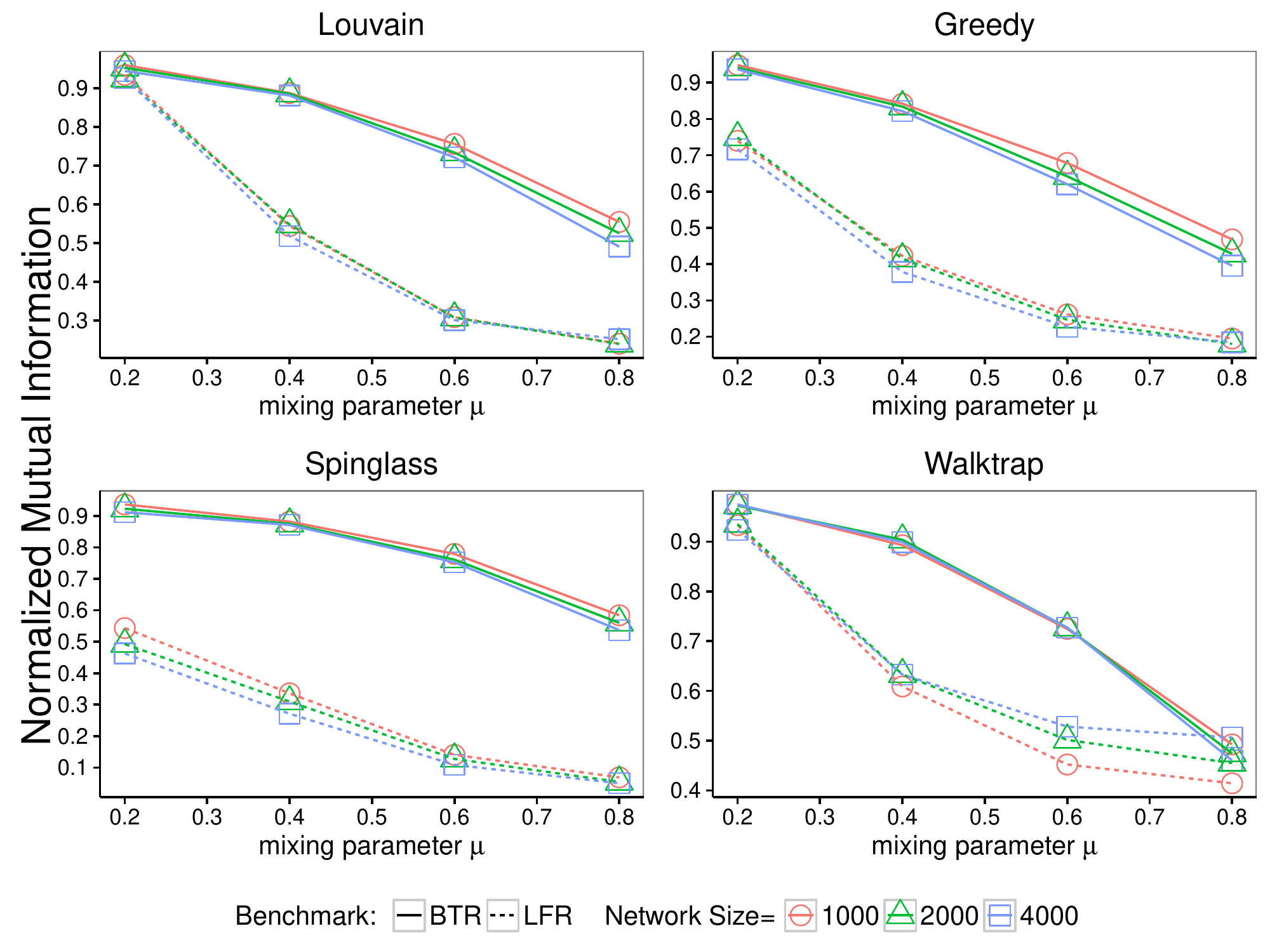}
\label{fig::comp_ind-sim}
\end{subfigure}
\begin{subfigure}[t]{0.48\textwidth}
\includegraphics[width=\textwidth]{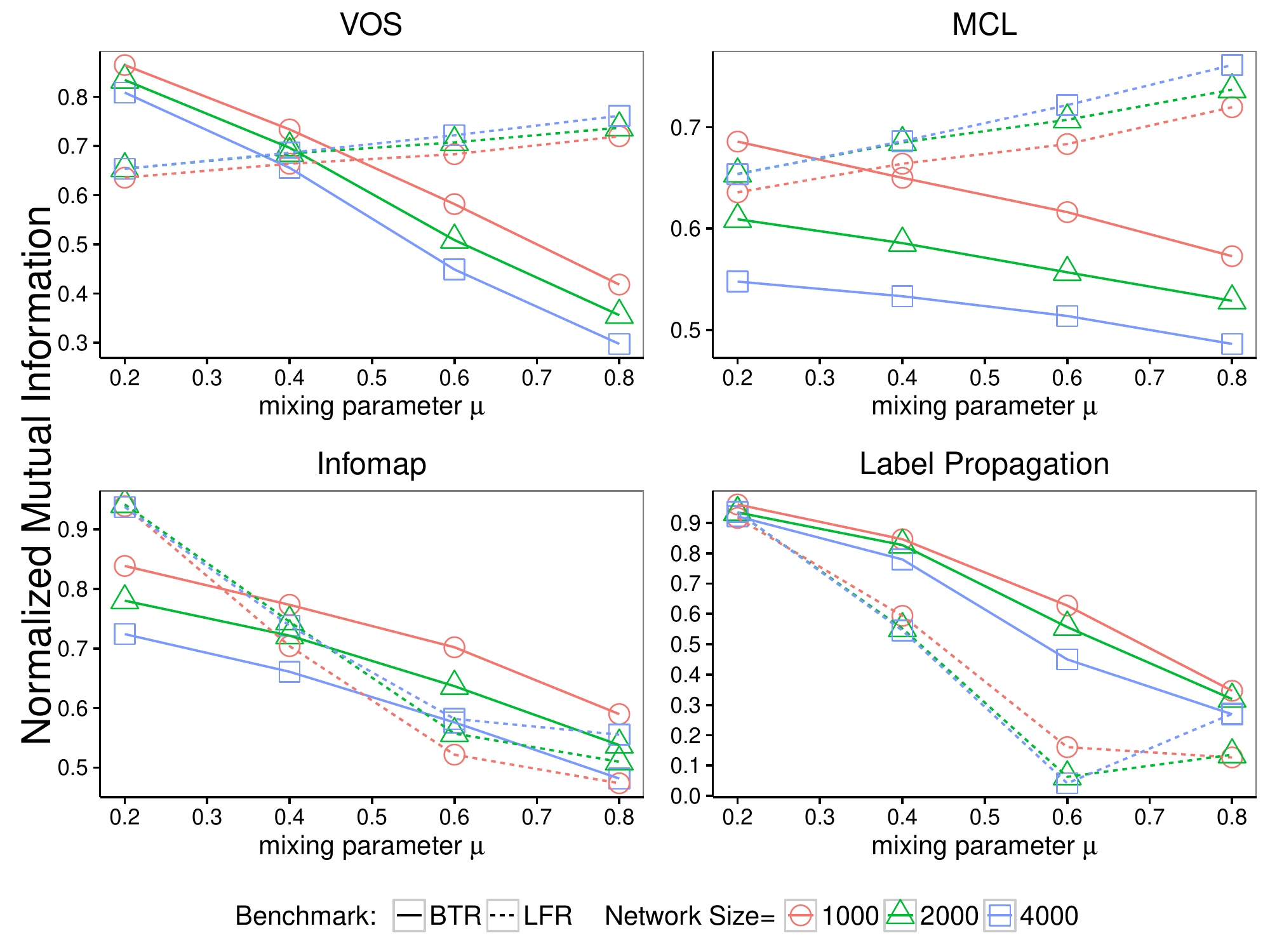}
\label{fig::comp_ind-diff}
\end{subfigure}
\caption{Performance of eight algorithms on the two benchmarks. Each plot represents one community detection algorithm for which results are averaged over all instances. Solid lines shows results on graphs generated by LFR benchmark whereas dotted line shows the results for graphs generated by the proposed benchmark.}
\label{fig::compind}
\end{figure}

\textbf{Figure \ref{fig::compind}:}
\begin{enumerate}
\item Comparison of Community Detection Algorithms
\item Performance of eight algorithms on the two benchmarks. Each plot represents one community detection algorithm for which results are averaged over all instances. Solid lines shows results on graphs generated by LFR benchmark whereas dotted line shows the results for graphs generated by the proposed benchmark.
\end{enumerate}


\end{document}